\documentclass[%
 reprint,
 amsmath,amssymb,
 aps,
aps,prx,
]{revtex4-2} %linenumbers,

\usepackage{graphicx} % Include figure files
\usepackage{dcolumn}  % Align table columns on decimal point
\usepackage{bm}       % bold math
\usepackage{verbatim} %add comments
\usepackage{gensymb}
\usepackage{lineno}   % numbers of rows
\usepackage{soul}     %
\usepackage{color}
\usepackage{textcomp}
\usepackage{CJK}
\usepackage{multirow}
\usepackage[super]{nth}

\usepackage[breaklinks]{hyperref}

\begin{document}
%\nolinenumbers

\preprint{APS/123-QED}

\title{Intercoupling of bulk acoustic streaming and acoustothermal effect: A high-frequency focused beam example}% Force line breaks with \\
%Static holographic acoustical tweezers for dynamic manipulation: Focused vortex

\author{Shiyu Li}

\affiliation{State Key Laboratory of Ocean Engineering, School of Ocean and Civil Engineering, Shanghai Jiao Tong University, Shanghai, 200240, China}

\author{Zhixiong Gong}
\email{zhixiong.gong@sjtu.edu.cn}
\affiliation{State Key Laboratory of Ocean Engineering, School of Ocean and Civil Engineering, Shanghai Jiao Tong University, Shanghai, 200240, China}
\affiliation{Key Laboratory of Marine Intelligent Equipment and System, Ministry of Education, China}

\date{\today}% It is always \today, today,

\begin{abstract}
High-frequency focused acoustic beams are promising for selective trapping of cells in fluids, but the related acoustic absorption may generate large acoustothermal effect which could cause thermal heating on cells or microparticles and bring extra acoustic body force due to the thermal gradient. 
The theory of the bulk acoustic streaming and acoustic radiation force in a focused-beam for the three-dimensional selective trapping of a cell has been developed [Li and Gong, Phys. Rev. Fluids, 11, 054201 (2026)], however, the acoustothermal effect and its feedback on the acoustic field at high frequency with strong absorption remain weakly understood. 
To solve this issue, we develop a theoretical and numerical model that couples acoustic propagation, bulk acoustic streaming, and acoustothermal effect in water. 
The acoustic body force is decomposed into a viscous-attenuation-induced acoustic body force $\mathbf{f}_{\mathrm{E}}$ and a temperature-gradient-induced acoustic body force $\mathbf{f}_{\mathrm{T}}$, while the temperature field is fed back to the frequency-domain acoustic calculation through the temperature-dependent material properties. 
Taking the single focused beam for example, within the pressure range constrained by the mechanical index, $\mathbf{f}_{\mathrm{T}}$ remains weaker than $\mathbf{f}_{\mathrm{E}}$, whereas streaming-induced convection can markedly reduce the temperature rise when the thermal Peclet number ($Pe_T$) exceeds unity. 
This work establishes a theoretical basis for predicting and controlling the intercoupling of bulk acoustic streaming and acoustothermal effec of high-frequency focused beams which will be helpful for the design of single-beam acoustical tweezers.

\end{abstract}

\pacs{Valid PACS appear here}% PACS, the Physics and Astronomy
%\keywords{Suggested keywords}%Use showkeys class option if keyword display desired
\maketitle

%------------------------------------------------------------------------------------------------
\section{\label{sec:Introduction}Introduction}

High-frequency focused ultrasound concentrates acoustic energy within a small focal region, enabling biomedical applications such as microparticle manipulation~\cite{baudoin2020acoustic,li2026review} and noninvasive therapy~\cite{ter2001high}. In the surrounding fluid, acoustic attenuation simultaneously generates a
time-averaged body force and localized heat generation: the former drives acoustic streaming, whereas the latter produces acoustothermal effect. These two processes are mutually coupled. Acoustic streaming redistributes the heat generated by acoustic absorption through convection, thereby modifying the temperature field, the updated temperature field changes the local material properties and feeds back on acoustic propagation, which in turn alters both the acoustic body force and the acoustic heat source.

Since the seminal works on bulk acoustic streaming, the theoretical framework has been systematically developed~\cite{eckart1948vortices,nyborg1953acoustic,westervelt1953theory,lighthill1978acoustic}. More recent reformulations have decomposed the acoustic Reynolds stress force into a conservative acoustic-Lagrangian gradient, which can be absorbed into the mean pressure, and a non-conservative source term that drives incompressible streaming~\cite{riaud2017influence}. This distinction is particularly important in thermally inhomogeneous focused beams, where temperature-dependent density and compressibility introduce additional temperature-gradient-induced acoustic body force~\cite{karlsen2016acoustic} beyond the conventional the viscous-attenuation-induced acoustic body force (associated with Eckart bulk acoustic streaming)~\cite{riaud2017influence}.

Previous ultrasound heating have shown that acoustic absorption can generate localized temperature rises and that streaming-induced convection can reshape the temperature field~\cite{wu1994effect,sheu2011acoustics,solovchuk2012simulation,solovchuk2013computational}. Meanwhile, studies of inhomogeneous and thermoviscous acoustofluidics have demonstrated that gradients of density and compressibility can directly generate acoustic body forces, and that acoustically induced temperature gradients may therefore alter streaming itself~\cite{karlsen2016acoustic,qiu2021fast,joergensen2023theory,joergensen2023transition}. These results indicate that, in high-frequency focused beams, acoustic streaming and heating should not be treated as one-way consequences of a prescribed acoustic field, but as parts of a coupled feedback loop involving absorption, convection, temperature-dependent material properties, and acoustic propagation.

Despite these advances, a self-consistent coupled model for high-frequency finite-aperture focused beams in free space remains lacking. Unlike confined standing wave microchannels, where thermoviscous boundary layers play central roles, free-space focused beams are governed primarily by bulk absorption, Eckart streaming, and convective heat redistribution in the focal region, requiring a dedicated coupling framework. Here, we develop such a model by deriving time-averaged governing equations from the conservation laws, decomposing the acoustic body force into the viscous-attenuation-induced acoustic body force
$\mathbf{f}_{\mathrm{E}}$ and the temperature-gradient-induced acoustic body force $\mathbf{f}_{\mathrm{T}}$, and evaluating the heat source using the expression of Das and Bhethanabotla~\cite{das2025acoustothermal}. The acoustic, streaming, and thermal fields are solved self-consistently through an outer iteration. The model is then used to quantify how acoustic pressure and frequency control the maximum temperature rise and streaming velocity, thereby clarifying the feedback among acoustic absorption, streaming convection, and thermally modified acoustic propagation.

The structure of this work is organized into four parts as follows. 
Sec.\ref{sec2:Theory} derives the governing equations for the acoustics--streaming--thermal coupling in a focused acoustic beam system, with particular emphasis on the decomposition of the  body force. Sec.\ref{sec3:Numerical} presents the numerical method, where an outer iteration is employed to solve the coupled acoustic, acoustic streaming, and temperature fields.
Sec.\ref{sec4: Results} presents the parametric study of acoustic pressure and frequency, and theoretically analyzes the rationality of the simulation results. Sec.\ref{sec5:conclusion} summarizes the main conclusions of the paper.

%------------------------------------------------------------------------------------------------------
%------------------------------------------------------------------------------------------------------
%------------------------------------------------------------------------------------------------------
\section{\label{sec2:Theory} Theoretical model}

\begin{figure} 
\centering 
\includegraphics[width=8.6cm]{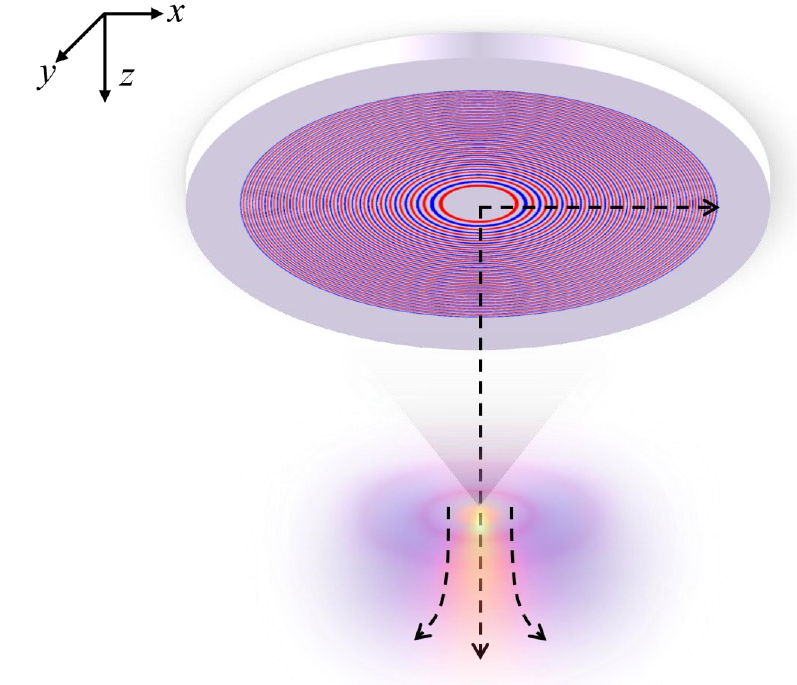}
\caption{
Schematic of the acoustics--streaming--thermal coupling system for a focused beam generated by a planar holographic transducer. Red and blue concentric electrodes denote two interleaved electrode sets driven with a \(\pi\)-phase difference. The focused acoustic field propagates along the axial direction and concentrates near the focal region, where viscous acoustic dissipation induces a localized temperature rise. The dashed arrows represent the time-averaged acoustic streaming, which transports heat away from the focal region. This schematic illustrates the coupled mechanism among the focused acoustic field (the light blue color), streaming flow (the arrows), and temperature field (the orange color).
}
\label{Fig1:Sche}
\end{figure}

We start from the conservation equations for a compressible Newtonian fluid. The mass and momentum conservation equations
are
\begin{equation}
\frac{\partial \rho}{\partial t}
+
\nabla\cdot(\rho\mathbf{v})
=0 ,
\label{eq:mass_conservation}
\end{equation}
and
\begin{equation}
\frac{\partial(\rho\mathbf{v})}{\partial t}
+
\nabla\cdot(\rho\mathbf{v}\mathbf{v})
=
-\nabla p
+
\nabla\cdot\boldsymbol{\sigma},
\label{eq:momentum_conservation}
\end{equation}
where $\rho$, $\mathbf{v}$, and $p$ denote the density, velocity, and
pressure, respectively. The viscous stress tensor is
\begin{equation}
\boldsymbol{\sigma}
=
\mu_s
\left[
\nabla\mathbf{v}
+
(\nabla\mathbf{v})^T
\right]
+
\left(
\mu_b-\frac{2}{3}\mu_s
\right)
(\nabla\cdot\mathbf{v})\boldsymbol{\mathsf{I}},
\label{eq:viscous_stress}
\end{equation}
where $\mu_s$ and $\mu_b$ are the dynamic and bulk viscosities, respectively. {$\mathsf{I}$ is the unit matrix.

The corresponding energy equation is written as Ref~\cite{bruus2007theoretical}
\begin{equation}
\begin{aligned}
\frac{\partial(\rho E)}{\partial t}
+
\nabla\cdot(\rho E\mathbf{v})
&=
-\nabla\cdot(p\mathbf{v})
+
\nabla\cdot(\boldsymbol{\sigma}\cdot\mathbf{v})  \\
&\quad
+
\nabla\cdot(k_{th}\nabla T),
\end{aligned}
\label{eq:energy_conservation}
\end{equation}
where $E=e+\frac{1}{2}|\mathbf{v}|^2$ 
is the total specific energy, $e$ is the internal energy per unit mass, $T$ is
the temperature, and $k_{th}$ is the thermal conductivity. Equation
\eqref{eq:energy_conservation} provides the energetic basis for the time-averaged heat-transfer equation used below, in which the irreversible loss of acoustic energy appears as the volumetric acoustic heat source.

In the present problem, the quantities of interest on the slow time scale ($\tau$) are the mean streaming flow and the mean temperature field, while the rapidly oscillating acoustic field (with the fast time scale $t$) provides the time-averaged momentum and energy
sources that drive the slow thermo-hydrodynamic response. We therefore use a time-scale decomposition rather than assigning all mean fields to a fixed order in a regular perturbation expansion. This formulation is
particularly suitable for acoustothermal problems~\cite{das2025acoustothermal}, because the steady temperature rise may exceed the oscillatory acoustic temperature perturbation and can feed back into the material properties and acoustic propagation. Each physical field variable is thus decomposed into a slowly varying mean component $\bar{g}(\mathbf{x},\tau)$ and a
rapidly oscillating acoustic component $\tilde{g}(\mathbf{x},t;\tau)$~\cite{joergensen2023theory},
\begin{equation}
g(\mathbf{x},t)
=
\bar{g}(\mathbf{x},\tau)
+
\tilde{g}(\mathbf{x},t;\tau),
\qquad
\langle \tilde{g}\rangle=0 ,
\label{eq:multiscale_decomposition}
\end{equation}
where $\langle\cdot\rangle$ denotes the
average over one acoustic period. 

The acoustic pressure and velocity are written in terms of their
complex amplitudes as
\begin{equation}
\tilde{p}
=
\mathrm{Re}
\left[
\hat{p}(\mathbf{x})e^{-i\omega t}
\right],
\qquad
\tilde{\mathbf{v}}
=
\mathrm{Re}
\left[
\hat{\mathbf{v}}(\mathbf{x})e^{-i\omega t}
\right].
\label{eq:harmonic_fields}
\end{equation}
Here, tildes denote fast oscillatory acoustic fields, whereas hats denote their complex amplitudes. Under the weak-damping approximation, the acoustic pressure field satisfies~\cite{joergensen2023theory,das2025acoustothermal}
\begin{equation}
\nabla^2 \hat{p}=-(\omega/c_w)^2(1+i\Gamma)\hat p++
O(\Gamma^2)
\label{eq:pressure_acoustics}
\end{equation}
where
\begin{equation}
\Gamma=\omega\bar{\kappa}\eta_L,
\qquad
\eta_L=\frac{4}{3}\bar{\mu}_s+\bar{\mu}_b,
\qquad
\Gamma\ll1 .
\label{eq:damping_factor}
\end{equation}
and $\bar{\kappa}=1/(\bar{\rho}\bar{c}^{2})$ is the steady compressibility. The associated acoustic velocity is
\begin{equation}
\hat{\mathbf{v}}
=
-\frac{i+\Gamma}{\omega\bar{\rho}}
\nabla \hat{p}.
\label{eq:acoustic_velocity}
\end{equation}
Note that this paper only considers bulk attenuation and does not take into account the influence of the boundary layer. The time-averaged acoustic intensity is defined as
\begin{equation}
\mathbf{I}_{\mathrm{ac}}
=
\left\langle
\tilde{p}\tilde{\mathbf{v}}
\right\rangle
=
\frac{1}{2}
\mathrm{Re}
\left[
\hat{p}\hat{\mathbf{v}}^*
\right],
\label{eq:acoustic_intensity}
\end{equation}
where the superscript $*$ denotes complex conjugation.

The slow mean flow is driven by the time-averaged body force generated by the first-order acoustic field. To leading order in the acoustic Mach number, the acoustic body force can be expressed as
\begin{equation}
\mathbf{F}_{\mathrm{ac}}
=
-\nabla\cdot
\left\langle
\bar{\rho}
\tilde{\mathbf{v}}\otimes\tilde{\mathbf{v}}
\right\rangle
=
-\frac{1}{2}
\mathrm{Re}
\left\{
\nabla\cdot
\left[
\bar{\rho}
\hat{\mathbf{v}}\otimes\hat{\mathbf{v}}^*
\right]
\right\}.
\label{eq:reynolds_force}
\end{equation}
For weak damping and slowly varying material properties, this force can be
decomposed as
\begin{equation}
\mathbf{F}_{\mathrm{ac}}
=
\nabla\mathcal{L}_{\mathrm{ac}}
+
\mathbf{f}_{\mathrm{ac}},
\label{eq:body_force_decomposition}
\end{equation}
where
\begin{equation}
\mathcal{L}_{\mathrm{ac}}
=
\frac{1}{4}\bar{\kappa}|\hat{p}|^2
-
\frac{1}{4}\bar{\rho}|\hat{\mathbf{v}}|^2
\label{eq:acoustic_lagrangian}
\end{equation}
is the acoustic Lagrangian density, and
\begin{equation}
\mathbf{f}_{\mathrm{ac}}
=
-\frac{1}{4}|\hat{p}|^2\nabla\bar{\kappa}
-\frac{1}{4}|\hat{\mathbf{v}}|^2\nabla\bar{\rho}
+
\frac{\Gamma\omega}{\bar{c}_w^{2}}\mathbf{I}_{\mathrm{ac}} .
\label{eq:effective_acoustic_force}
\end{equation}
The detailed derivation of this decomposition is provided in Appendix~\ref{Appendix A}. The conservative term $\nabla\mathcal{L}_{\mathrm{ac}}$ can be absorbed into the mean pressure~\cite{riaud2017influence}. Therefore, $\mathbf{f}_{\mathrm{ac}}$ is the effective
non-conservative body force that drives the incompressible streaming flow. The first two terms in Eq.~\eqref{eq:effective_acoustic_force} arise from
temperature-induced gradients of density and compressibility and are denoted by
\(\mathbf{f}_{\mathrm{T}}\), whereas the last term is the viscous-attenuation-induced acoustic body force \(\mathbf{f}_{\mathrm{E}}\). This expression can recover the established limiting cases. For a thermally homogeneous but attenuating fluid,
\(\nabla\bar{\rho}=\nabla\bar{\kappa}=0\), and the body force reduces to
the \(\mathbf{f}_{\mathrm{ac}}=\mathbf{f}_{\mathrm{E}}\)~\cite{riaud2017influence,li2024eckart}.
Conversely, in the nondissipative inhomogeneous limit with \(\Gamma=0\), it reduces
to the acoustic force density associated with density and compressibility
gradients, \(\mathbf{f}_{\mathrm{ac}}=\mathbf{f}_{\mathrm{T}}\)~\cite{qiu2021fast}. The present
formulation therefore retains both contributions simultaneously and extends
these limiting descriptions to thermally coupled focused beams.

The steady streaming flow is governed by
\begin{equation}
\nabla\cdot\bar{\mathbf{u}}=0 ,
\label{eq:streaming_continuity}
\end{equation}
\begin{equation}
\bar{\rho}
(\bar{\mathbf{u}}\cdot\nabla)\bar{\mathbf{u}}
=
-\nabla\Pi
+
\nabla\cdot
\left[
{\bar \mu_s}
\left(
\nabla\bar{\mathbf{u}}
+
\nabla\bar{\mathbf{u}}^{T}
\right)
\right]
+
\mathbf{f}_{\mathrm{ac}},
\label{eq:streaming_momentum}
\end{equation}
where $\bar{\mathbf{u}}$ is the time-averaged streaming velocity and $\Pi$ is the reduced mean pressure after absorbing the conservative acoustic contribution (see Eq.~\eqref{eq:PI}). Note that the average momentum equation only takes into account the changes in physical properties $\bar\mu_s$ caused by the mean temperature field.

 The steady time-averaged heat-transfer equation is written as
\begin{equation}
\bar{\rho}\bar{c}_p
\bar{\mathbf{u}}_{\mathrm{L}}\cdot\nabla\bar{T}
=
\nabla\cdot
\left(
\bar{k}_{th}\nabla\bar{T}
\right)
+
Q_{\mathrm{ac}},
\label{eq:heat_transfer}
\end{equation}
where $\bar c_p$ is specific heat capacity at constant pressure, and $\bar{\mathbf{u}}_{\mathrm{L}}$ denotes the Lagrangian mean velocity,
or equivalently the mass-transport velocity, defined as
\begin{equation}
\bar{\mathbf{u}}_{\mathrm{L}}
=
\bar{\mathbf{u}}
+
\frac{
\left\langle
\tilde{\rho}\tilde{\mathbf{v}}
\right\rangle
}{\bar{\rho}} .
\label{eq:lagrangian_velocity}
\end{equation}
Here, $\bar{\mathbf{u}}$ is the Eulerian mean streaming velocity, and the second term represents the acoustic mass-flux velocity contribution. When this contribution is neglected in the heat-advection term, the Lagrangian mean velocity is approximated by the Eulerian mean velocity,
$\bar{\mathbf{u}}_{\mathrm{L}}\simeq\bar{\mathbf{u}}$.

In the present work, the acoustic heat source is evaluated using the acoustothermal heat-source expression proposed in
Ref.~\cite{das2025acoustothermal}, rather than the local plane-wave absorption
approximation. From the acoustic energy balance, the heat source can be related to the divergence of the acoustic Poynting vector,
\begin{equation}
Q_{\mathrm{ac}}
=
-\nabla\cdot
\left\langle
\tilde{p}\tilde{\mathbf{v}}
\right\rangle .
\label{eq:poynting_heat_source}
\end{equation}
Under the weak-damping approximation, this expression reduces to the following form
\begin{equation}
Q_{\mathrm{ac}}
=
\Gamma\bar{\rho}\omega
\left\langle
\tilde{\mathbf{v}}\cdot\tilde{\mathbf{v}}
\right\rangle .
\label{eq:closed_heat_time_average}
\end{equation}
Using the complex velocity amplitude in Eq.~\eqref{eq:harmonic_fields}, the time average becomes
\begin{equation}
\left\langle
\tilde{\mathbf{v}}\cdot\tilde{\mathbf{v}}
\right\rangle
=
\frac{1}{2}
|\hat{\mathbf{v}}|^2 ,
\end{equation}
and therefore the volumetric heat source used in the model is
\begin{equation}
Q_{\mathrm{ac}}
=
\frac{1}{2}
\Gamma\bar{\rho}\omega
|\hat{\mathbf{v}}|^2 .
\label{eq:closed_heat_source}
\end{equation}
For a locally plane traveling wave, Eq.~\eqref{eq:closed_heat_source} recovers
the conventional expression $Q_{\mathrm{ac}}\simeq
2\alpha|\mathbf{I}_{\mathrm{ac}}|$, with
$\alpha=\Gamma\omega/(2{\bar c_w})$, and $\bar c_w$ is the steady sound speed. Thus, the plane-wave heat source is only a
limiting form of the expression, while
Eq.~\eqref{eq:closed_heat_source} is used in the present simulations.

Equations~\eqref{eq:pressure_acoustics}, \eqref{eq:streaming_continuity},\eqref{eq:streaming_momentum}, and
\eqref{eq:heat_transfer}, together with the acoustic force
\eqref{eq:effective_acoustic_force} and heat source
\eqref{eq:closed_heat_source}, constitute the theoretical framework for the coupled acoustic propagation, bulk streaming, and heat transport.
 The mean temperature field changes the local material properties, thereby modifying acoustic propagation and, consequently, the acoustic body force and heat source. Conversely, acoustic absorption generates heat, and the streaming flow redistributes this heat by convection, which updates the temperature field. It should be noted that we do not take into account the influence of acoustic streaming on sound propagation.
For clarity, the governing components of the closed model are summarized as
\begin{equation} \begin{aligned} &\text{Acoustic field:} \\ &\qquad \nabla^{2}\hat{p} = -\left(\frac{\omega}{c_w}\right)^2 (1+i\Gamma)\hat{p} +O(\Gamma^2), \\[0.8em] &\text{Streaming field:} \\ &\qquad \nabla\cdot\bar{\mathbf{u}}=0, \\ &\qquad \bar{\rho} (\bar{\mathbf{u}}\cdot\nabla)\bar{\mathbf{u}} = -\nabla\Pi + \nabla\cdot \left[ \bar{\mu}_s \left( \nabla\bar{\mathbf{u}} + \nabla\bar{\mathbf{u}}^{T} \right) \right] + \mathbf{f}_{\mathrm{ac}}, \\[0.8em] &\text{Temperature field:} \\ &\qquad \bar{\rho}\bar{c}_p \bar{\mathbf{u}}_{\mathrm{L}}\cdot\nabla\bar{T} = \nabla\cdot \left( \bar{k}_{\mathrm{th}}\nabla\bar{T} \right) + Q_{\mathrm{ac}}, \\[0.8em] &\text{Acoustic body force:} \\ &\qquad \mathbf{f}_{\mathrm{ac}} = -\frac{1}{4} |\hat{p}|^{2}\nabla\bar{\kappa} -\frac{1}{4} |\hat{\mathbf{v}}|^{2}\nabla\bar{\rho} + \frac{\Gamma\omega}{\bar{c}_w^{2}} \mathbf{I}_{\mathrm{ac}}, \\[0.8em] &\text{Acoustic heat source:} \\ &\qquad Q_{\mathrm{ac}} = \frac{1}{2} \Gamma\bar{\rho}\omega |\hat{\mathbf{v}}|^{2}. \end{aligned} \label{eq:model_summary} \end{equation}

\section{Numerical implementation \label{sec3:Numerical}}

Considering the symmetry of the system, a two-dimensional axisymmetric model was adopted to reduce the computational cost. Owing to the clear separation between the acoustic oscillation time scale and the hydrodynamic/thermal time scales, the coupling model was resolved by an outer iteration. The acoustic field was treated as a first-order time-harmonic field, whereas the streaming flow and the temperature field were solved on the slow time scale. The simulations were performed using the commercial finite-element software COMSOL Multiphysics 6.3.
The first-order acoustic field was computed using the \textit{Pressure Acoustics, Frequency Domain} interface, while the streaming flow and temperature fields were solved as steady problems using the \textit{Laminar Flow} and \textit{Heat Transfer in Fluids} interfaces, respectively. The outer iteration among the acoustic, acoustic streaming, and thermal models were controlled through COMSOL Multiphysics 6.3 with MATLAB. The computational mesh was discretized with a maximum element size of $\lambda/16$~\cite{li2026competition}.

The coupling among the acoustic, streaming, and thermal fields was implemented
through an outer iteration. An initial temperature field
\(T^{(0)}\), taken as the uniform ambient temperature, was first prescribed.
At the \(m\)th outer iteration, the temperature field \(T^{(m)}\) was used to update the temperature-dependent material properties. For \(m=0\), \(T^{(m)}\) corresponds to the prescribed initial temperature field, whereas for \(m\geq 1\), it is the temperature field obtained from the preceding steady streaming and heat-transfer solve. The frequency-domain first-order acoustic field,
\(p_1^{(m)}\) and \(\mathbf{v}_1^{(m)}\), was then computed, from which the
acoustic heat source \(Q_{\mathrm{ac}}^{(m)}\) and the time-averaged acoustic
body force \(\mathbf{f}_{\mathrm{ac}}^{(m)}\) were evaluated.
During each outer iteration, the acoustic source terms were kept fixed and
supplied to the steady flow and heat-transfer equations, yielding the updated
streaming velocity \(\mathbf{u}^{(m+1)}\) and temperature field \(T^{(m+1)}\). The updated temperature field was directly fed back into the
next acoustic solve without applying any relaxation. The overall iteration
procedure is illustrated in Fig.~\ref{Fig2: Simulation workflow}.

\begin{figure} 
\centering 
\includegraphics[width=8.6cm]{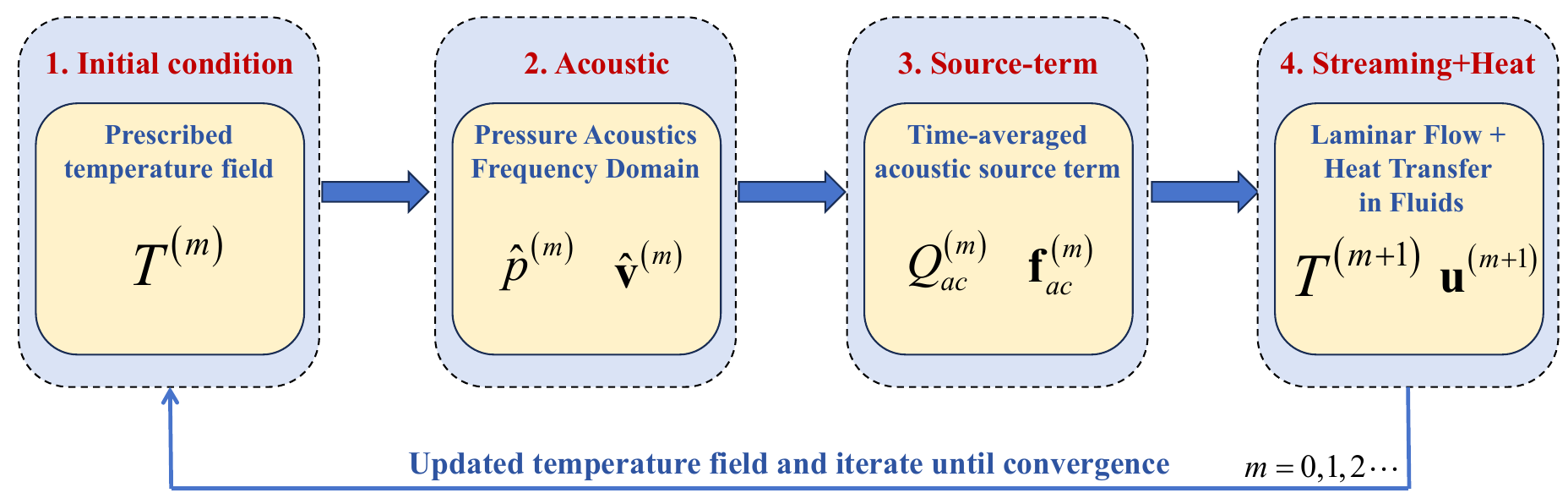}
\caption{
Schematic of the outer iteration for the temperature-feedback model. At iteration \(m\), the acoustic field is solved in the frequency domain using the current temperature field \(T^{(m)}\). The resulting acoustic heat source \(Q_{\mathrm{ac}}^{(m)}\) and streaming body force \(\mathbf{f}_{\mathrm{ac}}^{(m)}\) are then imposed in the coupled Laminar Flow and Heat Transfer in Fluids problem to obtain \(\mathbf{u}^{(m+1)}\) and \(T^{(m+1)}\). The updated temperature field is fed back into the acoustic model, and the loop is repeated until convergence.
}

\label{Fig2: Simulation workflow}
\end{figure}

The outer iteration was continued until the relative changes in the maximum temperature rise and maximum streaming velocity were both smaller than
$10^{-4}$:
\begin{equation}
\varepsilon_T^{(m)}
=
\frac{
\left|
\Delta T_{\max}^{(m+1)}
-
\Delta T_{\max}^{(m)}
\right|
}{
\Delta T_{\max}^{(m+1)}
}
<10^{-4},
\end{equation}
and
\begin{equation}
\varepsilon_u^{(m)}
=
\frac{
\left|
u_{\max}^{(m+1)}
-
u_{\max}^{(m)}
\right|
}{
u_{\max}^{(m+1)}
}
<10^{-4}.
\end{equation}
All acoustic, thermal, and streaming quantities reported in this work were extracted after convergence of the outer iteration. This procedure retains
the time-averaged driving effect of the acoustic field on the slow hydrodynamic and temperature fields, while allowing the temperature-induced variation of material properties to feed back into the acoustic calculation.
It therefore captures the nonlinear coupling among acoustic absorption, streaming-induced convective transport, and thermal feedback in a high-frequency focused acoustic beam.

%-----Sec IV
\section{\label{sec4: Results} Results and Discussion}

\subsection{Physical parameters and field distributions}

We first specify the focused beam configuration and the material parameters used in the simulations. The acoustic field was generated by a planar holographic transducer consisting of two interleaved sets of concentric electrodes driven with a phase difference of \(\pi\) as illustrate in Fig.\ref{Fig1:Sche}. For a design frequency \(f_0\), the acoustic wavenumber in water is defined as \(k_0=2\pi f_0/c_{\mathrm{w}}\), where \(c_{\mathrm{w}}\) is the sound speed in water. The radii of the two electrode sets are given by~\cite{gong2022single,li2025reversing}
\begin{eqnarray} & & R_1 = \frac{1}{k_0} \sqrt{(C+2 n \pi)^2 - (k_0h_0)^2} \label{circ1} \\ & & R_2 = \frac{1}{k_0} \sqrt{[C+ (2 n+1) \pi]^2 - (k_0h_0)^2}
\label{circ2} 
\end{eqnarray}
where \(h_0\) is the design focal length, \(n\) is the electrode index, and \(C\) is a constant. In the reference configuration, the transducer was designed with \(f_0=5~\mathrm{MHz}\), \(h_0=6~\mathrm{mm}\), and \(N=17\), with \(n\in[0,N]\). Once \(f_0\), \(h_0\), and \(N\) are specified, the aperture radius \(R_{\mathrm{ap}}\) is determined by the electrode geometry.
For the reference case considered below, the focal pressure amplitude is normalized to \(p_f=1~\mathrm{MPa}\). This normalization allows the streaming velocity and temperature rise to be compared under the same local acoustic forcing level.

The surrounding fluid is water as shown in Table.\ref{Table 1 Acoustic properties}. Its density, sound speed, dynamic and bulk viscosities, thermal conductivity, specific heat capacity, and acoustic attenuation are treated as temperature-dependent material properties in the coupled simulations. The initial and ambient temperature is set to \(T_0=293.15~\mathrm{K}\,(20^\circ\mathrm{C})\). Unless otherwise stated, the transducer boundary is treated as thermally insulating so that no additional heat is supplied by the transducer itself, while the remaining outer boundaries are maintained at \(T_0\). This choice isolates the temperature rise generated by acoustic viscous dissipation inside the fluid domain.

Fig.~\ref{Fig3:Field} shows the representative field distributions for the reference configuration. In Fig.~\ref{Fig3:Field}(a), the left half presents the focused acoustic pressure field, while the right half shows the acoustic-induced temperature field calculated without convective heat transport. The localized pressure focus produces acoustic heating near the focal region, with the maximum temperature increasing to \(293.168~\mathrm{K}\) (\(20.018^\circ\mathrm{C}\)). This result is referred to as the no-convection temperature field.
Figure~\ref{Fig3:Field}(b) shows the coupled streaming and temperature fields. The left half gives the acoustic streaming field, whereas the right half gives the temperature field obtained when streaming-induced heat convection is included. Compared with the no-convection case, the coupled temperature field exhibits a lower peak value, with the maximum temperature increasing only to $293.156~\mathrm{K}$ ($20.006^\circ\mathrm{C}$), indicating that acoustic streaming redistributes the generated heat and suppresses thermal accumulation near the focus. This comparison highlights the role of the mean streaming flow in modifying the acoustic-induced temperature field.

\begin{figure} 
\centering 
\includegraphics[width=8.6cm]{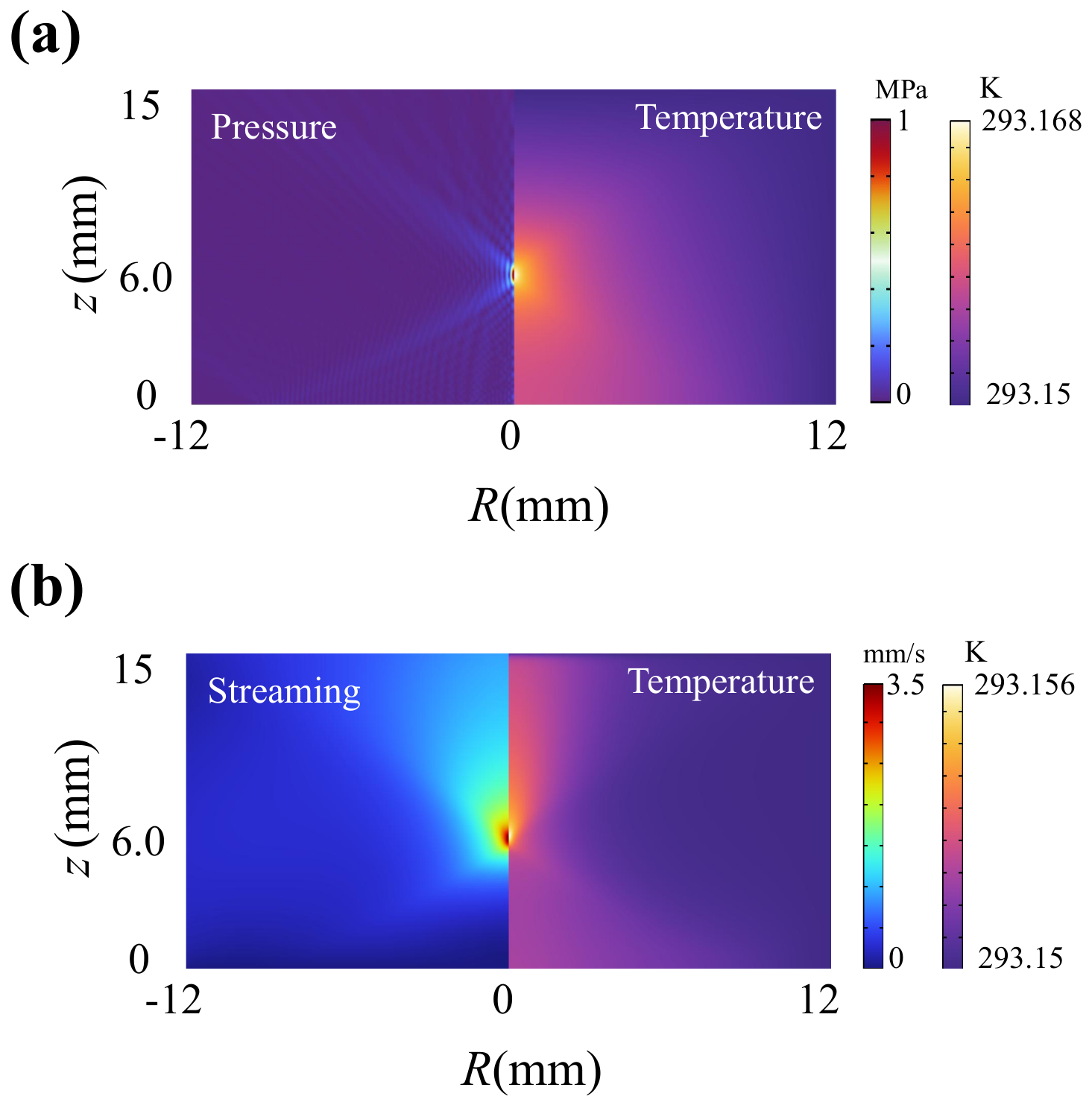}
\caption{
Acoustic, streaming, and temperature field distributions for the reference case with \(p_f=1~\mathrm{MPa}\), \(N=17\), \(f_0=5~\mathrm{MHz}\), and \(h_0=6~\mathrm{mm}\). 
(a) Acoustic pressure field and temperature field predicted from acoustic heating alone, without flow convection. 
(b) Coupled streaming velocity and temperature fields obtained by solving the steady streaming and heat-transfer equations.
The coupled result shows that acoustic streaming transports heat away from the focal region and lowers the peak temperature compared with the purely conductive temperature field.
}
\label{Fig3:Field}
\end{figure}

\begin{table}[!htbp]
\small
  \caption{ Physical parameters at $20^\circ C$ of the water. Density, sound speed, heat capacity, and thermal conductivity were taken from the NIST thermophysical-property database, and the dynamic viscosity follows the IAPWS formulation. The bulk viscosity was estimated
as \(\mu_b=2.79\mu_s\), following the COMSOL water material definition.} 
  \label{Table 1 Acoustic properties}
  \begin{tabular*}{0.48\textwidth}{@{\extracolsep{\fill}}lll}
    \hline 
    Symbol & Physical parameter      & Value \\
    \hline 
    $\rho$   & Mass density      & 998 kg/m$^3$    \\
    $c_w$      & Sound velocity   & 1482 m/s   \\
    $c_p$      & Speciﬁc heat capacity   & 4182 $Jkg^{-1}K^{-1}$   \\
    $k_{th}$   &  Thermal conductivity   & 0.60~$Wm^{-1}K^{-1}$    \\
    $\mu_s$    & dynamic viscosity   & $1.002 \times 10^{-3}$ Pa s  \\
    $\mu_b$    & bulk viscosity      & $2.80 \times 10^{-3} $ Pa s \\
    $\eta_L$      & longitudinal viscosit & $\frac{4}{3}{\mu}_s+{\mu}_b$    \\
    \hline
  \end{tabular*}
\end{table}

\subsection{Effect of acoustic pressure on temperature rise and streaming velocity}

The acoustic pressure range was chosen with reference to the mechanical index (MI), a dimensionless output index commonly used in diagnostic ultrasound to assess the likelihood of cavitation-related nonthermal mechanical bioeffects.
Following Apfel and Holland~\cite{apfel1991gauging}, the MI is defined as
\begin{equation}
    \mathrm{MI}=\frac{p_{r,\max}}{\sqrt{f_c}},
\end{equation}
where \(p_{r,\max}\) is the derated peak rarefactional pressure in MPa and \(f_c\) is the center frequency in MHz. The commonly adopted upper limit \(\mathrm{MI}\leq 1.9\) was used as a reference safety criterion. For the present focused beam operating near \(5~\mathrm{MHz}\), this limit corresponds to
\begin{equation}
    p_{r,\max} \lesssim 1.9\sqrt{5}\simeq 4.25~\mathrm{MPa}.
\end{equation}
Therefore, the focal acoustic pressure considered in this study was restricted
to \(p_f\leq 4~\mathrm{MPa}\). Within this range, we examine how increasing acoustic pressure modifies the temperature rise, acoustic streaming velocity, and their mutual coupling. Under these conditions, the coupled temperature rise remains sufficiently small, so that a single outer iteration is sufficient to satisfy the prescribed convergence criteria.

Figure~\ref{Fig4:Pressure} shows the pressure dependence of the maximum streaming velocity and maximum temperature rise. The no-convection temperature rise, \(\Delta \bar T_0\), increases approximately quadratically with the focal pressure. This behavior follows directly from the acoustic heat source,
\begin{equation}
    Q_{\mathrm{ac}} \sim p_f^2,
\end{equation}
and the conduction-dominated balance
\begin{equation}
    \bar k_{th}\frac{\Delta \bar T_0}{L_T^2}\sim Q_{\mathrm{ac}},
\end{equation}
which gives \(\Delta \bar T_0\sim p_f^2\) when the thermal length scale \(L_T\) is fixed. In contrast, the coupled temperature rise, \(\Delta \bar T_c\), deviates from this quadratic trend as the pressure increases. This deviation is caused by streaming-induced convective heat transport. The relative importance of convection is measured by the thermal Peclet number,
\begin{equation}
    Pe_T=\frac{U L_T}{\bar\chi},
    \label{eq:Pe}
\end{equation}
where \(U\) is the characteristic streaming velocity, \(L_T\) is the thermal length scale, and \(\bar\chi=\bar k_{th}/(\bar\rho \bar c_p)\) is the thermal diffusivity. Taking \(L_T\simeq\lambda\) as a first estimate gives \(Pe_T=1\) at approximately \(p_f\simeq0.35~\mathrm{MPa}\). This threshold is consistent with the observed separation between \(\Delta \bar T_0\) and \(\Delta \bar T_c\): above this pressure, streaming-induced convection becomes strong enough to transport heat away from the focal region, and the coupled temperature rise grows more slowly than the purely conductive prediction.

The pressure-dependent scaling of Eckart streaming in focused beams has been systematically investigated by Li and Gong~\cite{li2026competition}. Here, we briefly recall the key scaling arguments relevant to the present results. In the low-Reynolds-number limit, the mean flow is governed by the Stokes balance, which gives
\begin{equation}
    U\sim p_f^2.
\end{equation}
This explains the quadratic pressure dependence of the creeping flow solution.

As \(p_f\) increases, the hydrodynamic Reynolds number,
\begin{equation}
    Re_\lambda=\frac{U L_U}{\nu},
\end{equation}
approaches unity. Here, \(U\) is the characteristic streaming velocity, \(L_U\) is the characteristic length scale of the streaming flow, and \(\nu=\mu_s/\rho\) is the kinematic viscosity of the fluid. Taking \(L_U=\lambda\) as the characteristic length gives \(Re_\lambda=1\) at approximately \(p_f\simeq1~\mathrm{MPa}\). Above this threshold, the inertial term becomes non-negligible. As shown by Li and Gong~\cite{li2026competition}, inertial transport in a focused streaming jet reduces the pressure exponent from the Stokes-limit value $n=2$ toward the inertial-limit scaling $n=4/3$. Therefore, the deviation of the laminar flow velocity from the creeping flow prediction is mainly attributed to hydrodynamic inertia, rather than to the temperature-gradient body force.

The coupled velocity remains very close to the Eckart streaming velocity over the entire pressure range considered. This indicates that, under the present conditions, the streaming is primarily governed by the viscous-attenuation-induced body force, whereas the temperature-gradient-induced acoustic body force provides only a weak correction.
The comparison can be made using the
compressibility-gradient contribution,
\begin{equation}
    \mathbf{f}_{T}
    \simeq
    -\frac{1}{4}|\hat p|^2\nabla\bar\kappa ,
    \label{eq:thermal_force_kappa}
\end{equation}
where \(\bar\kappa=1/(\bar\rho\bar c_w^2)\). The density-gradient term
is not retained in this estimate, because for water near room
temperature, the compressibility-gradient contribution is the dominant
temperature-gradient contribution~\cite{joergensen2023theory}. Using
\begin{equation}
|\nabla\bar{\kappa}|
\sim
\left|
\frac{\partial\bar{\kappa}}{\partial T}
\right|
\frac{\Delta\bar{T}_{c}}{L_{T}},
\label{eq:kappa_gradient_estimate}
\end{equation}
the temperature-gradient body force scales as
\begin{equation}
f_{T}
\sim
\frac{|\hat{p}|^{2}}{4}
\left|
\frac{\partial\bar{\kappa}}{\partial T}
\right|
\frac{\Delta\bar{T}_{c}}{L_{T}} .
\label{eq:fT_scaling}
\end{equation}
Here \(L_T\) is the effective length scale over which the coupled temperature field varies.

The corresponding viscous-attenuation-induced body force is estimated as
\begin{equation}
f_{E}
\sim
\frac{\Gamma\omega}{\bar{c}_{w}^{2}} I_{\rm ac}
\sim
\frac{\Gamma\omega}{\bar{c}_{w}^{2}}
\frac{|\hat{p}|^{2}}{\bar{\rho}\bar{c}_{w}},
\label{eq:fE_scaling}
\end{equation}
where a local travelling-wave estimate has been used. Therefore, the relative-force ratio is
\begin{equation}
\Lambda_T
=
\frac{f_T}{f_E}
\sim
\frac{\bar{\rho}\bar{c}_{w}^{3}}{4\Gamma\omega}
\left|
\frac{\partial\bar{\kappa}}{\partial \bar T}
\right|
\frac{\Delta\bar{T}_{c}}{L_T}.
\label{eq:LambdaT_1}
\end{equation}
Equivalently, using
\begin{equation}
\Gamma
=
\frac{\omega\eta_L}{\bar{\rho}\bar{c}_{w}^{2}},
\qquad
\eta_L=\frac{4}{3}\bar{\mu}_s+\bar{\mu}_b ,
\end{equation}
Eq.~\eqref{eq:LambdaT_1} becomes
\begin{equation}
\Lambda_T
\sim
\frac{\bar{\rho}^{2}\bar{c}_{w}^{5}}{4\omega^{2}\eta_L}
\left|
\frac{\partial\bar{\kappa}}{\partial \bar T}
\right|
\frac{\Delta\bar{T}_{c}}{L_T}.
\label{eq:LambdaT_2}
\end{equation}
For water at \(25^{\circ}{\rm C}\) and \(f=5~{\rm MHz}\), using
\(\left|\partial\bar{\kappa}/\partial T\right|\simeq 1.0\times10^{-12}~{\rm Pa^{-1}K^{-1}}\), Eq.~\eqref{eq:LambdaT_2} gives
\begin{equation}
\Lambda_T
\sim
1.5\times10^{-3}
\Delta\bar{T}_{c}[{\rm mK}]
\left(\frac{\lambda}{L_T}\right),
\label{eq:LambdaT_numeric}
\end{equation}
where \(\lambda=\bar{c}_{w}/f\).

This magnitude estimate is conservative because \(L_T\) is not generally equal to the acoustic wavelength. In the present focused beam, the Eckart streaming advects and spreads the heated region, so the coupled temperature field is smoothed over a length scale comparable to, or larger than, the acoustic wavelength. Thus \(L_T\gtrsim\lambda\), and the temperature-gradient contribution is weaker than the estimate obtained by setting \(L_T=\lambda\). For example, at \(p_0=4~{\rm MPa}\), the no-convection temperature rise is about \(281~{\rm mK}\), in the coupled case, convection suppresses and spreads the temperature field, giving a maximum temperature rise of only about \(29~{\rm mK}\). The same conservative estimate then gives \(\Lambda_T\sim0.04\), and the actual value is smaller if \(L_T>\lambda\).

Beyond this magnitude comparison, the vortical content of the body force is crucial. For the temperature-gradient force, Eq.~\eqref{eq:thermal_force_kappa} may be written as
\begin{equation}
\mathbf{f}_{T}
\simeq
-A_T(\mathbf{x})\nabla\bar{T},
\qquad
A_T(\mathbf{x})
=
\frac{1}{4}|\hat{p}|^{2}
\frac{\partial\bar{\kappa}}{\partial \bar T}.
\end{equation}
Its curl is therefore
\begin{equation}
\nabla\times\mathbf{f}_{T}
=
-\nabla A_T\times\nabla\bar{T}
\simeq
-\frac{1}{4}
\frac{\partial\bar{\kappa}}{\partial T}
\left(
\nabla|\hat{p}|^{2}\times\nabla\bar{T}
\right).
\label{eq:curl_fT}
\end{equation}
 In the present free-space focused beam, the temperature field is mainly generated by acoustic absorption and is therefore spatially correlated with the acoustic intensity field. As a result, the curl of \(\mathbf{f}_{T}\) is small. This explains why the coupled velocity remains close to the Eckart streaming prediction even though a finite temperature gradient is present.

The temperature field is nevertheless strongly affected by the mean flow, because even a weakly modified velocity field can produce significant convective heat transport once \(Pe_T\gtrsim1\). Thus, the pressure response is characterized by two distinct transitions: the temperature field first departs from the conduction-dominated quadratic scaling when \(Pe_T\) becomes order unity, whereas the streaming velocity departs from the creeping flow quadratic scaling only later, when \(Re_\lambda\) approaches unity.
\begin{figure} 
\centering 
\includegraphics[width=8.6cm]{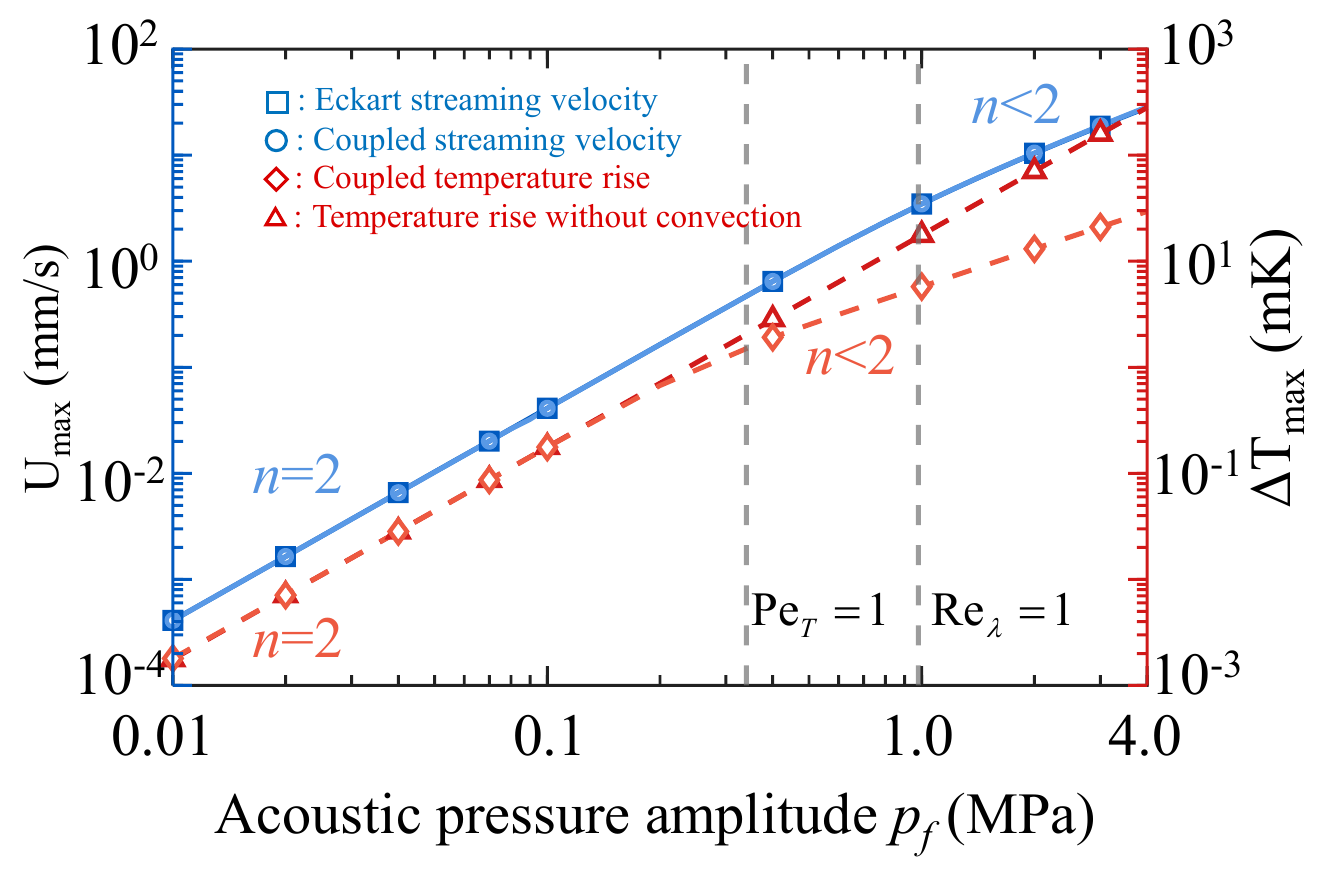}
\caption{
Log--log pressure dependence of the maximum streaming velocity and temperature
rise for \(p_f=0.01\)--\(4~\mathrm{MPa}\) with fixed transducer geometry. The
left ordinate shows the Eckart streaming velocity \(U_E\) and coupled streaming velocity \(U_c\),
and the right ordinate shows the conductive temperature rise
\(\Delta\bar{T}_0\) and coupled temperature rise \(\Delta\bar{T}_c\). Dashed
vertical lines mark the estimated \(Pe_T=1\) and \(Re_\lambda=1\) thresholds.
Because \(\Lambda_T\ll1\), \(U_c\) nearly overlaps with \(U_E\), at
\(p_f=4~\mathrm{MPa}\), \(U_c=2.847~\mathrm{cm/s}\) and
\(U_E=2.811~\mathrm{cm/s}\). The low-pressure regime follows the quadratic
scaling \(n=2\), while \(Pe_T\sim1\) indicates convective suppression of
temperature rise and \(Re_\lambda\sim1\) indicates inertial deviation of the
streaming velocity from creeping-flow scaling.
}
\label{Fig4:Pressure}
\end{figure}

\subsection{Effect of frequency on temperature rise and streaming velocity}

Because the maximum acoustic pressure allowed by the MI constraint depends on frequency, a fixed focal pressure of $p_f=1~\mathrm{MPa}$ was used in the frequency study. This choice ensures that the acoustic pressure remains within the safety limit over the entire frequency range considered.

Figure~\ref{Fig5:Frequency} shows the frequency dependence of the maximum streaming velocity and maximum temperature rise at a fixed focal pressure. Two streaming velocities are compared: the Eckart streaming velocity, obtained by retaining only the attenuation-induced body force ($\mathbf{f}_{\mathrm{E}}$), and the coupled velocity, obtained by including temperature-dependent feedback in the steady streaming and heat-transfer solve. The thermal response is evaluated in the same manner by comparing the conductive temperature rise, calculated without flow convection, with the coupled temperature rise when streaming-induced convective heat transport is included.

\begin{figure} 
\centering 
\includegraphics[width=8.6cm]{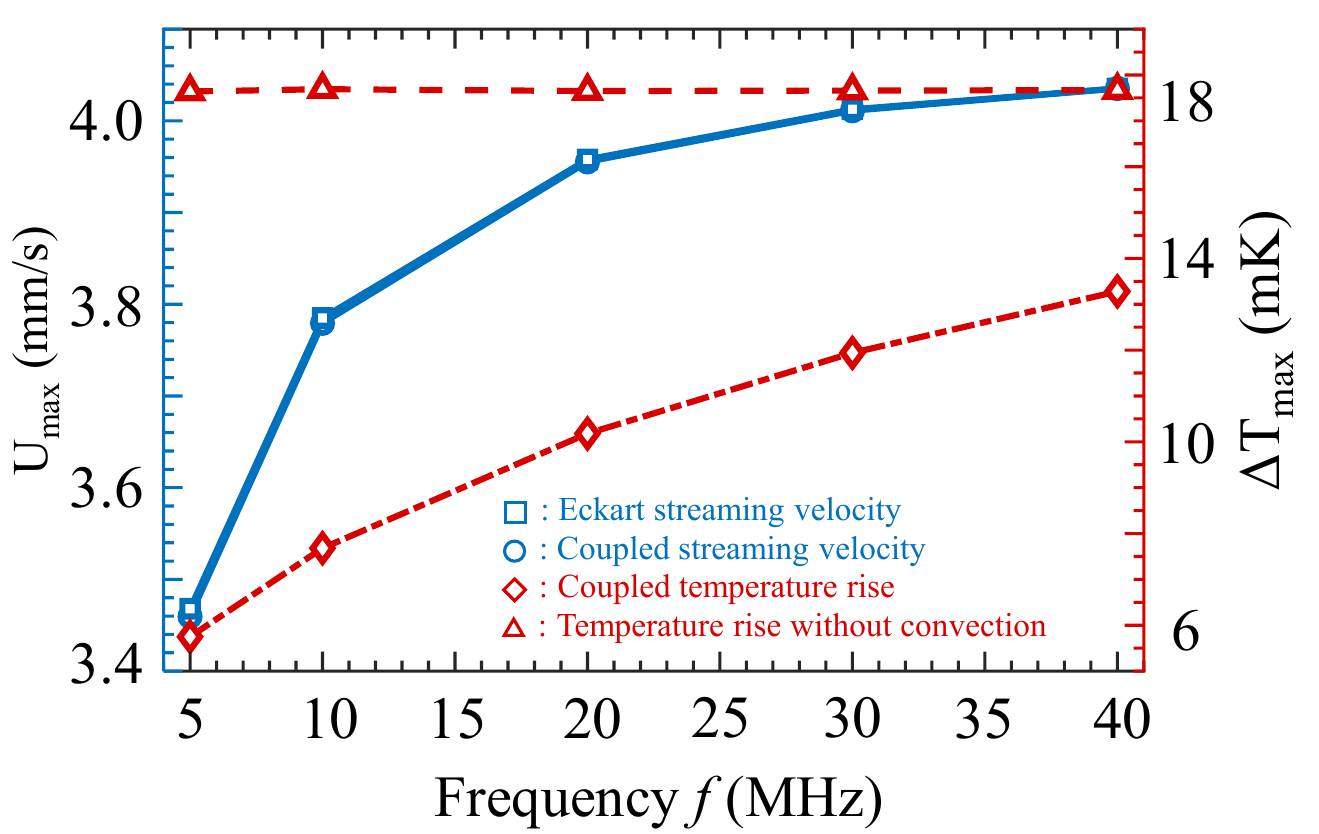}
\caption{
Frequency dependence of the maximum streaming velocity and maximum temperature rise. The focal pressure was fixed at \(p_f=1~\mathrm{MPa}\), the number of electrode turns was \(N=17\). The transducer geometry was redesigned at each frequency to maintain a constant aperture angle. Blue curves: Eckart and coupled streaming velocities. Red curves: temperature rise without convection and coupled temperature rise. The velocity exhibits a weak frequency dependence, whereas the coupled temperature rise increases monotonically with frequency.
}
\label{Fig5:Frequency}
\end{figure}

The conductive temperature rise remains nearly invariant over the investigated frequency range. This behavior can be understood from a simple scaling argument. At fixed focal pressure, the local acoustic velocity amplitude near the focus is approximately fixed, whereas the viscous acoustic heat source scales as
\begin{equation}
    Q_{\mathrm{ac}} \sim \omega^2 |\mathbf{v}_1|^2 \sim f^2 .
\end{equation}
However, under the present focusing configuration, the characteristic thermal source length decreases with the wavelength,
\begin{equation}
    L_T \sim \lambda \sim f^{-1}.
\end{equation}
In the conduction-dominated limit,
\begin{equation}
    \Delta \bar T_{0}
    \sim
    \frac{Q_{\mathrm{ac}} L_T^2}{\bar k_{th}}
    \sim
    f^2 f^{-2}
    \sim
    f^0 .
\end{equation}
Thus, the frequency-induced enhancement of local viscous dissipation is largely compensated by the reduction of the heated region. The nearly flat no-convection temperature curve is therefore consistent with the expected scaling for geometrically similar focused heating.

The streaming velocity also exhibits only a weak frequency dependence. The Eckart streaming maximum velocity increases from \(3.5~\mathrm{mm\,s^{-1}}\) at \(5~\mathrm{MHz}\) to \(4.0~\mathrm{mm\,s^{-1}}\) at \(40~\mathrm{MHz}\). The coupled velocity almost overlaps with the Eckart streaming velocity over the entire frequency range. This indicates that, at \(p_f=1~\mathrm{MPa}\), the mean flow is still dominated by $\mathbf{f}_{E}$, while $\mathbf{f}_{T}$ only provides a weak correction.

The weak frequency dependence of the Eckart velocity follows from the same compensation mechanism. The $\mathbf{f}_{E}$ scales as
\begin{equation}
    \mathbf{f}_{E} \sim \frac{\alpha \mathbf{I}_{\mathrm{ac}}}{\bar c_w},
\end{equation}
where $\alpha=\frac{\Gamma\omega}{2\bar{c}} \sim f^2$ is the attenuation coefficient. Since the focal pressure, and therefore the local acoustic intensity $\mathbf{I}_{\mathrm{ac}}$ is fixed, one obtains \(f_E \sim f^2\). In the Stokes-dominated limit, the streaming velocity scales as
\begin{equation}
    U_E \sim \frac{f_E L_U^2}{\bar\mu_s}.
\end{equation}
With \(L_U \sim f^{-1}\), the velocity scale becomes
\begin{equation}
    U_E \sim f^2 f^{-2} \sim f^0 .
\end{equation}
This explains why the velocity does not follow the \(f^2\) scaling suggested by the attenuation coefficient alone. Similar weak frequency dependence has also been reported for high-frequency focused acoustic tweezers, where the growth of attenuation is counterbalanced by the reduction of the acoustic source dimensions with wavelength~\cite{vincent2026high}.

In contrast to the nearly invariant conductive temperature rise, the coupled temperature rise increases monotonically from \(5.75\) to \(13.28~\mathrm{mK}\). This increase is not caused by a larger conductive heating level, but rather by the weakening of convective heat removal at higher frequencies. The trend can be interpreted using the thermal Peclet number defined in Eq.~\eqref{eq:Pe}. Because the maximum streaming velocity $U$ varies only weakly with frequency while the characteristic thermal length scale decreases approximately as $L_T\sim f^{-1}$, $Pe_T$ decreases as the frequency increases. Thus, streaming-induced heat convection becomes less effective at higher frequencies, and the coupled temperature rise progressively approaches the conduction-dominated value.

These results show that the frequency response of the coupled thermoacoustic system is governed by the interplay between enhanced viscous dissipation and the reduction of the characteristic focal-region length scale.
 The maximum streaming velocity is nearly frequency independent because the increase in ${f}_{E}$ is compensated by the reduction of the wavelength-scaled acoustic forcing region. By contrast, the coupled temperature rise increases with frequency because the reduced thermal length scale suppresses convective heat transport. Therefore, even though the velocity and temperature fields originate from the same acoustic dissipation process, they exhibit distinct frequency dependencies once thermal convection is included.

%-----------------------Sec V
\section{\label{sec5:conclusion} Conclusions}

In this work, we developed a theoretical and numerical framework to describe the intercoupling between bulk acoustic streaming and acoustothermal effect, and demonstrated it using a high-frequency focused beam in water as a representative case. Starting from the time-averaged momentum equation, the acoustic body force was decomposed into the viscous-attenuation-induced acoustic body force $\mathbf{f}_{E}$ and an additional temperature-gradient-induced acoustic body force $\mathbf{f}_{T}$.
An outer-iteration strategy was then constructed to solve the coupled problem. At each iteration, the frequency-domain acoustic field was first computed using the current temperature-dependent material properties. The resulting acoustic heat source and streaming body force were then supplied to the slow-time streaming and heat-transfer problem to update the mean velocity and temperature fields. This framework allows the feedback from temperature-dependent material properties and streaming-induced heat convection to be incorporated in a controlled manner.

The pressure-parametric study revealed two distinct transitions. At low pressure, the temperature rise and streaming velocity follow the expected quadratic scaling with the focal pressure, consistent with the \(p_f^2\) dependence of the acoustic heat source and the acoustic body force. As the pressure increases, the thermal Peclet number first becomes of order unity, leading to a clear departure of the coupled temperature rise $\Delta \bar T_{c}$ from the purely conductive prediction $\Delta \bar T_{0}$ due to streaming-induced convective heat transport. At higher pressure, the streaming Reynolds number approaches unity, and hydrodynamic inertia causes the laminar flow velocity to deviate from the creeping flow scaling. In contrast, the coupled velocity remains close to the Eckart laminar velocity, because the estimated $\mathbf{f}_{T}$ remains smaller than the $\mathbf{f}_{E} $ over the investigated pressure range, and has only a weak vortical component.

The frequency-parametric study showed that, under fixed focal pressure and geometrically similar focusing, both the maximum streaming velocity and the purely conductive temperature rise exhibit only weak frequency dependence. This behavior results from the compensation between enhanced viscous dissipation and the reduced characteristic size of the focal heat-source region at higher frequencies. When streaming-induced heat convection is included, however, the
coupled temperature rise increases with frequency because the decreasing thermal length scale lowers the thermal Peclet number and weakens convective heat removal.

Overall, the present results advance the modeling of realistic focused acoustic tweezers by incorporating acoustothermal effect, acoustic streaming, and temperature feedback within a unified framework. The analysis identifies the relevant pressure and frequency scaling laws and provides practical guidance for choosing operating parameters. These findings provide a basis for future studies of high-frequency focused-beam manipulation in thermally sensitive fluids and biological environments. Future work will leverage our transducer platform~\cite{li2026experimental}, together with the guidance provided by the present theoretical model, to explore advanced applications.

%-----------------------
\begin{acknowledgments}
Z. Gong thanks for the support from the National Natural Science Foundation of China (24Z990200542 and No. 12504522), the XIAOMI Foundation, and the Shanghai Jiao Tong University [2030 Initiative, AI for Engineering Initiative, and the startup funding (WH220401017, WH22040121)].
\end{acknowledgments}

%-----------------------
\appendix
\section{\label{Appendix A} Decomposition of the Reynolds stress force}

This appendix summarizes the derivation of the body force
decomposition used in the main text. The derivation starts from the time-averaged Reynolds stress and is carried out under the assumptions of small acoustic Mach number, weak acoustic damping, and slowly varying background material properties.

The first-order acoustic pressure and velocity are written as
\begin{equation}
\tilde{p}
=
\mathrm{Re}
\left[
\hat{p}e^{-i\omega t}
\right],
\qquad
\tilde{\mathbf{v}}
=
\mathrm{Re}
\left[
\hat{\mathbf{v}}e^{-i\omega t}
\right],
\label{eq:app_harmonic_fields}
\end{equation}
where hats denote complex amplitudes. For two time-harmonic quantities,
\(\tilde{a}=\mathrm{Re}[\hat{a}e^{-i\omega t}]\) and
\(\tilde{b}=\mathrm{Re}[\hat{b}e^{-i\omega t}]\), the time-average is
\begin{equation}
\left\langle
\tilde{a}\tilde{b}
\right\rangle
=
\frac{1}{2}
\mathrm{Re}
\left[
\hat{a}\hat{b}^{*}
\right],
\label{eq:app_average_identity}
\end{equation}
where the superscript \(^*\) denotes complex conjugation.

To leading order in the acoustic Mach number, the mean acoustic force density acting on the slow flow can be written in the Reynolds-stress form
\begin{equation}
\mathbf{F}_{\mathrm{ac}}
=
-\nabla\cdot
\left\langle
\bar{\rho}
\tilde{\mathbf{v}}\otimes\tilde{\mathbf{v}}
\right\rangle
=
-\frac{1}{2}
\mathrm{Re}
\left\{
\nabla\cdot
\left[
\bar{\rho}
\hat{\mathbf{v}}\otimes\hat{\mathbf{v}}^{*}
\right]
\right\}.
\label{eq:app_reynolds_force}
\end{equation}
Here, the overbar denotes a slowly varying mean quantity. The density, sound
speed, and compressibility are functions of the mean temperature field,
\(\bar{\rho}=\rho(\bar{T})\), \(\bar{c}=c(\bar{T})\), and
\(\bar{\kappa}=1/(\bar{\rho}\bar{c}^{2})\).

We next specify the weakly damped acoustic equations used to evaluate
Eq.~\eqref{eq:app_reynolds_force}. The linearized continuity equation and
equation of state give
\begin{equation}
-i\omega\bar{\kappa}\hat{p}
+
\nabla\cdot\hat{\mathbf{v}}
=
0 .
\label{eq:app_linear_continuity}
\end{equation}
For a longitudinal acoustic wave in a Newtonian fluid, the linearized momentum
equation with viscous damping is
\begin{equation}
-i\omega\bar{\rho}\hat{\mathbf{v}}
=
-\nabla\hat{p}
+
\eta_L
\nabla
\left(
\nabla\cdot\hat{\mathbf{v}}
\right),
\label{eq:app_linear_momentum}
\end{equation}
where
\begin{equation}
\eta_L
=
\frac{4}{3}\bar{\mu}_s
+
\bar{\mu}_b
\label{eq:app_longitudinal_viscosity}
\end{equation}
is the longitudinal viscosity. Substituting
Eq.~\eqref{eq:app_linear_continuity} into
Eq.~\eqref{eq:app_linear_momentum} and treating the material properties as
frozen over one acoustic wavelength gives
\begin{equation}
-i\omega\bar{\rho}\hat{\mathbf{v}}
=
-
\left(
1-i\Gamma
\right)
\nabla\hat{p},
\label{eq:app_momentum_damped}
\end{equation}
with
\begin{equation}
\Gamma
=
\omega\bar{\kappa}\eta_L,
\qquad
\Gamma\ll1 .
\label{eq:app_gamma}
\end{equation}
Therefore,
\begin{equation}
\hat{\mathbf{v}}
=
-\frac{i+\Gamma}{\omega\bar{\rho}}
\nabla\hat{p}
+
O(\Gamma^2).
\label{eq:app_velocity_pressure_relation}
\end{equation}

Substitution of Eq.~\eqref{eq:app_velocity_pressure_relation} into
Eq.~\eqref{eq:app_linear_continuity} yields the weakly damped pressure
acoustic equation
\begin{equation}
\nabla^2 \hat{p}=-k^2_0(1+i\Gamma)\hat p++
O(\Gamma^2)
\label{eq:app_pressure_operator}
\end{equation}
This is the pressure equation used in the theoretical model.

To make the decomposition explicit, we retain terms of $O(\Gamma)$. Using
the weakly damped pressure--velocity relation~\eqref{eq:expanded_reynolds_force_app}, Eq.~\eqref{eq:app_reynolds_force}
can be written as
\begin{equation}
\mathbf{F}_{\mathrm{ac}}
=
-\frac{1}{2\omega^{2}}
\mathrm{Re}
\left\{
\nabla\cdot
\left[
\frac{1}{\bar{\rho}}
\nabla\hat{p}\otimes\nabla\hat{p}^{*}
\right]
\right\}.
\label{eq:expanded_reynolds_force_app}
\end{equation}

Expanding the divergence gives
\begin{equation}
\mathbf{F}_{\mathrm{ac}}
=
-\frac{1}{2\omega^{2}}
\mathrm{Re}
\left[
\left(
\frac{1}{\bar{\rho}}\nabla\hat{p}^{*}\cdot\nabla
\right)
\nabla\hat{p}
+
\nabla\hat{p}\,
\nabla\cdot
\left(
\frac{1}{\bar{\rho}}\nabla\hat{p}^{*}
\right)
\right].
\label{eq:expanded_pressure_force_app}
\end{equation}

The first term is evaluated using
\begin{equation}
2\mathrm{Re}
\left[
(\nabla\hat{p}^{*}\cdot\nabla)\nabla\hat{p}
\right]
=
\nabla|\nabla\hat{p}|^{2},
\qquad
|\nabla\hat{p}|^{2}
=
\omega^{2}\bar{\rho}^{2}|\hat{\mathbf v}|^{2},
\label{eq:kinetic_identity_app}
\end{equation}
whereas the second term is reduced using the weakly damped pressure equation Eq.~\eqref{eq:app_pressure_operator}.

Keeping the retained-order terms yields
\begin{equation}
\mathbf{F}_{\mathrm{ac}}
=
\frac{1}{4}
\bar{\kappa}
\nabla|\hat{p}|^{2}
-
\frac{1}{4}
\bar{\rho}
\nabla|\hat{\mathbf v}|^{2}
-
\frac{1}{2}
|\hat{\mathbf v}|^{2}
\nabla\bar{\rho}
+
\frac{\Gamma\omega}{\bar{c}^{2}}
\mathbf{I}_{\mathrm{ac}},
\qquad
\label{eq:force_intermediate_app}
\end{equation}

Equation~\eqref{eq:force_intermediate_app}
is then rearranged by adding and subtracting the gradients of the slowly
varying material properties:
\begin{equation}
\frac{1}{4}
\bar{\kappa}
\nabla|\hat{p}|^{2}
=
\nabla
\left(
\frac{1}{4}
\bar{\kappa}|\hat{p}|^{2}
\right)
-
\frac{1}{4}
|\hat{p}|^{2}
\nabla\bar{\kappa},
\end{equation}
and
\begin{equation}
-\frac{1}{4}
\bar{\rho}
\nabla|\hat{\mathbf{v}}|^{2}
-
\frac{1}{2}
|\hat{\mathbf{v}}|^{2}
\nabla\bar{\rho}
=
-\nabla
\left(
\frac{1}{4}
\bar{\rho}|\hat{\mathbf{v}}|^{2}
\right)
-
\frac{1}{4}
|\hat{\mathbf{v}}|^{2}
\nabla\bar{\rho}.
\end{equation}
Substitution of these two identities into
Eq.~\eqref{eq:force_intermediate_app} yields
\begin{equation}
\mathbf{F}_{\mathrm{ac}}
=
\nabla \mathcal{L}_{\mathrm{ac}}
-
\frac{1}{4}
|\hat{\mathbf{v}}|^{2}
\nabla\bar{\rho}
-
\frac{1}{4}
|\hat{p}|^{2}
\nabla\bar{\kappa}
+
\frac{\Gamma\omega}{\bar{c}^{2}}
\mathbf{I}_{\mathrm{ac}},
\label{eq:force_decomposition_app}
\end{equation}
where
\begin{equation}
\mathcal{L}_{\mathrm{ac}}
=
\frac{1}{4}
\bar{\kappa}|\hat{p}|^{2}
-
\frac{1}{4}
\bar{\rho}|\hat{\mathbf{v}}|^{2}
\label{eq:acoustic_lagrangian_app}
\end{equation}
is the acoustic Lagrangian density. The first term in
Eq.~\eqref{eq:force_decomposition_app} is conservative and can be absorbed
into the mean pressure. The remaining terms constitute the effective
non-conservative acoustic body force driving the streaming flow.

Here,
\begin{equation}
\mathcal{L}_{\mathrm{ac}}
=
\frac{1}{4}
\bar{\kappa}
|\hat{p}|^{2}
-
\frac{1}{4}
\bar{\rho}
|\hat{\mathbf{v}}|^{2}
\label{eq:app_acoustic_lagrangian}
\end{equation}
is the acoustic Lagrangian density. Since
\(\nabla\mathcal{L}_{\mathrm{ac}}\) is conservative, it can be absorbed into
the mean pressure. Introducing the reduced mean pressure
\begin{equation}
\Pi
=
\bar{p}
-
\mathcal{L}_{\mathrm{ac}},
\label{eq:PI}
\end{equation}
the effective non-conservative body force that drives incompressible acoustic
streaming is
\begin{equation}
\mathbf{f}_{\mathrm{ac}}=\underbrace{-\frac{1}{4}|\hat{\boldsymbol{v}}|^2 \nabla \bar{\rho}-\frac{1}{4}|\hat{p}|^2 \nabla \bar{\kappa}}_{\mathbf{f}_{\mathrm{T}}}+\underbrace{\frac{\Gamma \omega}{\bar{c}^2} \boldsymbol{I}_{\mathrm{ac}}}_{\mathbf{f}_{\mathrm{E}}}
\label{eq:app_effective_force}
\end{equation}

The above equation agrees with Eq. (52)c of Ref.~\cite{joergensen2021theory} when setting the ratio of speciﬁc heat capacities $\gamma$ = 1 for the water case.
The first two terms in Eq.~\eqref{eq:app_effective_force} originate from
spatial variations of the mean density and compressibility,
\begin{equation}
\mathbf{f}_{\mathrm{T}}
=
-
\frac{1}{4}
|\hat{\mathbf{v}}|^{2}
\nabla\bar{\rho}
-
\frac{1}{4}
|\hat{p}|^{2}
\nabla\bar{\kappa}.
\label{eq:app_temperature_gradient_force}
\end{equation}
When the inhomogeneity is induced by the mean temperature field, this
contribution can be written as
\begin{equation}
\mathbf{f}_{\mathrm{T}}
=
-
\frac{1}{4}
|\hat{\mathbf{v}}|^{2}
\frac{\partial\bar{\rho}}{\partial\bar{T}}
\nabla\bar{T}
-
\frac{1}{4}
|\hat{p}|^{2}
\frac{\partial\bar{\kappa}}{\partial\bar{T}}
\nabla\bar{T}.
\label{eq:app_thermal_force}
\end{equation}
The above equation agrees with Eq. (1) of Ref.~\cite{qiu2021fast}.
This decomposition shows that, in a thermally inhomogeneous focused beam, the
mean flow is driven by both the viscous-attenuation-induced acoustic body force $\mathbf{f}_{\mathrm{E}}$
and the temperature-gradient-induced acoustic body force $\mathbf{f}_{\mathrm{T}}$.

\renewcommand\refname{Reference}
\bibliography{main}        %Produce the bibliography

\end{document}